\documentclass[12pt,a4paper,fleqn]{article}
\usepackage[DIV12]{typearea}
\usepackage{amsmath,amsfonts,amssymb}
\usepackage{graphicx}
\usepackage{cite}
\usepackage{mathrsfs}
\usepackage{bbm}
\usepackage{pifont}
\usepackage{units}
\usepackage{longtable}
\usepackage{hhline}
\usepackage[small,loose]{subfigure}  

\newcommand{\eVdist}{\kern-0.06em}

\newcommand{\rep}[1]{\ensuremath\boldsymbol{#1}}
\newcommand{\crep}[1]{\ensuremath\overline{\boldsymbol{#1}}}

\DeclareMathOperator{\tr}{tr}

\newcommand{\I}{\mathrm{i}}

\newcommand{\E}[1]{\ensuremath{\mathrm{E}_{#1}}} 

\newcommand{\SO}[1]{\ensuremath{\mathrm{SO}(#1)}}
\newcommand{\SU}[1]{\ensuremath{\mathrm{SU}(#1)}}
\newcommand{\U}[1]{\ensuremath{\mathrm{U}(#1)}}
\newcommand{\Z}[1]{\ensuremath{\mathbbm{Z}_{#1}}} 

\allowdisplaybreaks[1]

\title{A $\boldsymbol{\Z2\times\Z2}$ standard model}

\begin{document}

\begin{titlepage}

\begin{flushright}
TUM-HEP 741/09\\
LMU-ASC 53/09\\
HD-THEP-09-18
\end{flushright}


\begin{center}
{\Large\bf 
A $\boldsymbol{\Z2\times\Z2}$ standard model
}

\vspace{1cm}

\textbf{
Michael Blaszczyk\footnote[1]{Email: \texttt{michael@th.physik.uni-bonn.de}}{}$^a$,
Stefan Groot Nibbelink\footnote[2]{Email: \texttt{Groot.Nibbelink@physik.uni-muenchen.de}}{}$^{bc}$,
Michael Ratz\footnote[3]{Email: \texttt{mratz@ph.tum.de}}{}$^d$, 
Fabian Ruehle\footnote[4]{Email: \texttt{fabian@tphys.uni-heidelberg.de}}$^e$,
Michele Trapletti\footnote[5]{Email: \texttt{mtraplet@th.physik.uni-bonn.de}}{}$^f$,
Patrick K. S. Vaudrevange\footnote[6]{Email: \texttt{patrick.vaudrevange@physik.uni-muenchen.de}}{}$^b$
}
\\[5mm]
\textit{\small
{}$^a$ Bethe Center for Theoretical Physics and\\
Physikalisches Institut der Universit\"at Bonn,\\
Nussallee 12, 53115 Bonn, Germany
}
\\[3mm]
\textit{\small
{}$^b$ Arnold Sommerfeld Center for Theoretical Physics,\\
~~Ludwig-Maximilians-Universit\"at M\"unchen, 80333 M\"unchen, Germany
}
\\[3mm]
\textit{\small
{}$^c$ 
Shanghai Institute for Advanced Study, \\
University of Science and Technology of China,\\
99 Xiupu Rd, Pudong, Shanghai 201315, P.R.\ China
}
\\[3mm]
\textit{\small
{}$^d$ Physik-Department T30, Technische Universit\"at M\"unchen, \\
~~James-Franck-Stra\ss e, 85748 Garching, Germany
}
\\[3mm]
\textit{\small
{}$^e$ 
Institut f\"ur Theoretische Physik, Universit\"at Heidelberg, \\
Philosophenweg 16 und 19,  D-69120 Heidelberg, Germany
}
\\[3mm]
\textit{\small
{}$^f$ 
Institute of Theoretical Physics, Warsaw University, Ho\.za 69,
00-681, Warsaw, Poland
}

\end{center}

\vspace{1cm}

\begin{abstract}

We present a $\Z2\times\Z2$ orbifold compactification of the $\E8\times\E8$
heterotic string which gives rise to the exact chiral MSSM spectrum. The GUT
breaking $\SU5\to\SU3_C\times\SU2_\mathrm{L}\times\U1_Y$ is realized by modding
out a freely acting symmetry. This ensures precision gauge coupling unification. Further,
it allows us to break the GUT group without switching on flux in hypercharge
direction, such that the standard model gauge bosons can remain massless when
the orbifold singularities are blown up. The model has vacuum configurations
with matter parity, a large top Yukawa coupling and other phenomenologically
appealing features.

\end{abstract}

\end{titlepage}

\newpage

\section{Introduction}

Heterotic model building 
\cite{Dixon:1985jw,Dixon:1986jc,Ibanez:1986tp,Ibanez:1987sn,Casas:1987us}
has received renewed increased attention over the past few years. Almost
simultaneously, two constructions of models have been found that give rise to
the exact (chiral) spectrum of the minimal supersymmetric extension of the
standard model (SM), the MSSM. One of them is based on heterotic
\Z{6}-II orbifolds~\cite{Buchmuller:2005jr} and the other on smooth
Calabi-Yau (CY) compactifications~\cite{Bouchard:2005ag,Braun:2005nv}. (See
e.g.~\cite{Nilles:2008gq} for a review of recent progress in getting the MSSM from
string theory.)

Let us start by recalling some important properties of orbifold and CY
compactifications. Orbifolds are exact string compactifications in which one
directly goes from string theory in ten dimensions to an effective,
four-dimensional (4D) field theory. This ensures that one has an ultra-violet
complete framework in which couplings are accurately computable. The
mini-landscape of \Z6-II
orbifolds~\cite{Buchmuller:2006ik,Lebedev:2006kn,Lebedev:2007hv} provides a
large class of phenomenological appealing models of this kind: for example, the
MSSM matter spectrum is reproduced, vector-like exotics can be decoupled and
realistic features like a large top-Yukawa coupling, hierarchical couplings and
non-trivial flavor mixing emerge.

Orbifolds correspond to very special points in the string
landscape, where the worldsheet theory reduces to a combination of free
conformal field theories (CFTs). Generically, such orbifold models contain
unwanted gauge group factors and massless (vector-like) exotic states, which are
charged but not part of the standard model. In order to obtain
phenomenologically attractive vacua, non-trivial vacuum expectation values
(VEVs) need to be switched on. Furthermore, often non-vanishing VEVs cannot be
avoided due to the presence of a Fayet-Iliopoulos (FI) $D$-term for an anomalous
U(1). This means that the orbifold point almost never constitutes a true and
final vacuum configuration. Yet in all known examples ``nearby vacua'' can
be found in which some fields attain VEVs such as to cancel the FI term 
(see~\cite{Atick:1987gy}).

Like orbifolds, generic  compactifications of the heterotic string
that preserve $\mathcal{N}=1$ supersymmetry can give rise to models
with chiral spectra. They possess a large moduli space,
which may have special points like the large volume limit, the
orbifold and conifold points, etc. Each point in the moduli space
leads to an effective field theory with certain physical predictions,
like masses and couplings.
However, as the corresponding worldsheet theory involves a complicated
interacting CFT, most commonly only the supergravity limit, i.e.\ the lowest
order in $\alpha'$ approximation of the full string theory, is considered. In
this limit, generic compactifications have a clear geometrical interpretation
in terms of  CY manifolds, and the computation of the chiral spectra based
on index theorems is well under control. On the other hand, the validity of the
supergravity description requires moderately large radii, which is sometimes
problematic as this easily leads to too small gauge couplings. Moreover, since
the underlying CYs are complicated spaces, the calculation of couplings, needed
to make detailed predictions for phenomenology, is still far from
straightforward.

As is well known, orbifolds and CYs are not unrelated; rather, in many cases
orbifold singularities can be resolved, thus reproducing compactifications based
on smooth manifolds (see 
e.g.~\cite{Denef:2005mm,Lust:2006zh,GrootNibbelink:2007pn,GrootNibbelink:2007rd,Nibbelink:2008tv,Nibbelink:2009sp}). The transition from an orbifold to a smooth
compactification is achieved by giving VEVs to twisted states, stringy degrees
of freedom residing at the orbifold singularities. The reverse of the
``blow-up''
process, in which a compact hyper surface (i.e.\ exceptional
divisor) shrinks down to
zero size, is commonly referred to as a ``blow-down''.\footnote{Since the VEV and the corresponding volume are
schematically related by VEV $\sim$ exp(volume), the naive definition
of the volume has to go to $-\infty$ in order to arrive at the
orbifold point~\cite{Aspinwall:1993xz,Nibbelink:2009sp}. Hence, the blow-down in the supergravity sense does
not describe the orbifold point.} 
A setting in which one has both an exact orbifold CFT picture as well as a
smooth CY description, would be quite powerful, because one can combine the
calculability of the orbifold with the generic features of CY
compactifications.

So far no phenomenologically appealing model has been obtained that
allows for an orbifold as well as a CY description. It is unknown whether a
complete blow-down of the potentially realistic smooth compactifications,
obtained so far \cite{Bouchard:2005ag,Braun:2005nv,Anderson:2009mh}, to an exact
(free orbifold) CFT description exists. On the other hand, many
phenomenologically attractive orbifold
models~\cite{Buchmuller:2006ik,Lebedev:2006kn,Lebedev:2007hv} cannot be
completely blown up without destroying the phenomenological viability of these
settings, as the hypercharge or another part of the standard model gauge group
gets broken in the complete blow-up~\cite{Nibbelink:2009sp}. 

Our aim is to describe in detail how to construct (phenomenologically
attractive) orbifolds which allow for complete blow-ups without breaking the
standard model gauge group
$G_\mathrm{SM}=\SU3_C\times\SU2_\mathrm{L}\times\U1_Y$.
We base our discussion on an explicit model that can be seen as a specific
realization of a proposal made by Witten~\cite{Witten:1985xc} in that the GUT
breaking $\SU5\to G_\mathrm{SM}$ is achieved by dividing out a freely acting
symmetry.
(The idea of associating a Wilson line with an involution of the underlying CY
manifold in concrete model building has been employed already for some
time~\cite{Donagi:1999ez}.) This ensures that there is no flux in hypercharge
direction, such that $\U1_Y$ remains unbroken in the smooth limit. In addition,
such settings allow us to avoid GUT scale threshold corrections to the gauge
couplings and therefore fit particularly well to the paradigm of MSSM precision
gauge coupling unification~\cite{Hebecker:2004ce,Ross:2004mi}.

This letter is organized as follows: section~\ref{sec:model} is devoted to the
construction of a concrete orbifold model that allows for a freely acting
symmetry. The next section demonstrates that by switching on appropriate VEVs
quasi-realistic vacuum configurations can be obtained from our construction.
Section~\ref{sec:blowup} contains a tentative discussion of how to relate the
model to a smooth compactification in which all orbifold singularities have been
resolved. In section~\ref{sec:conclusions} we present our conclusions and
an outlook. Finally, the appendix contains the full spectrum of our model
and the string selection rules for allowed couplings.

\section{A $\boldsymbol{\Z2\times\Z2}$ orbifold model with a $\boldsymbol{\Z2}$ involution}
\label{sec:model}

We construct our model in two steps described in the following two subsections.
We start with a $\Z2\times\Z2$ orbifold based on the product of three two-tori
(for a detailed description of such orbifolds see \cite{Forste:2004ie} and
\cite{Cleaver:1998saa} for the free fermionic formulation) that leads to an \SU5
GUT with six generations and vector-like exotics.

In the second step we mod out a freely acting symmetry of order two. The
resulting geometry was first discussed in \cite{Donagi:2004ht} and corresponds
to model (1-1) in the classification by Donagi and
Wendland~\cite{Donagi:2008xy}. Since this $\Z2$ acts freely, the 48 fixed tori
of the $\Z2\times\Z2$ orbifold are mapped to each other pairwise resulting in 24
fixed tori. Hence, the number of chiral generations is reduced to three. The
final crucial ingredient of our model is a non-standard gauge embedding that
accompanies this involution and breaks \SU5 to $G_\mathrm{SM}$.

In the final two subsections we discuss some general phenomenological
properties of this model, like the massless spectrum and gauge coupling
unification.

\subsection{Underlying SU(5) $\boldsymbol{\Z2\times\Z2}$ orbifold}
\label{sec:orbifold}

The orbifold model is defined by a torus lattice that is spanned by six
orthogonal vectors $e_{\alpha}$, $\alpha=1,\ldots,6$, the $\Z2\times\Z2$ twist vectors $v_1 = (0,1/2,-1/2)$  and $v_2 =(-1/2,0,1/2)$, the  associated shifts 
\begin{subequations}
\begin{eqnarray}
 V_1 & = & \left( \frac{5}{4},-\frac{3}{4},-\frac{7}{4}, \frac{1}{4}, \frac{1}{4},-\frac{3}{4},-\frac{3}{4}, \frac{1}{4}\right)\, 
           \left( 0, 1, 1, 0, 1, 0, 0,-1 \right)\;, \\
 V_2 & = & \left(-\frac{1}{2},-\frac{1}{2},-\frac{1}{2}, \frac{1}{2},-\frac{1}{2},-\frac{1}{2},-\frac{1}{2},-\frac{1}{2}\right)\, 
           \left(\frac{1}{2}, \frac{1}{2}, 0, 0, 0, 0, 0, 4 \right)
\end{eqnarray}
\end{subequations}
and six discrete Wilson lines
\begin{subequations}
\begin{eqnarray}
W_1 & = & \left(  0^{8} \right)\,\left(  0^{8} \right)\;, \\
\label{eqn:W2}
W_2 & = & \left( \frac{5}{4}, \frac{1}{4}, \frac{3}{4},-\frac{1}{4},-\frac{1}{4}, \frac{3}{4}, \frac{3}{4}, \frac{3}{4} \right)\, 
          \left(-\frac{1}{4}, \frac{3}{4}, \frac{5}{4}, \frac{5}{4}, \frac{1}{4}, \frac{1}{4}, \frac{1}{4}, \frac{1}{4} \right)\;, \\
W_3 & = & \left(-\frac{3}{4},-\frac{1}{4}, \frac{1}{4}, \frac{7}{4},-\frac{1}{4},-\frac{1}{4},-\frac{1}{4},-\frac{1}{4} \right)\, 
          \left( \frac{1}{4}, \frac{1}{4}, \frac{1}{4}, \frac{5}{4},-\frac{3}{4}, \frac{1}{4},-\frac{3}{4}, \frac{1}{4} \right)\;, \\
W_5 & = & \left(-\frac{1}{2},-\frac{1}{2}, \frac{1}{2},-\frac{1}{2}, \frac{1}{2},-\frac{1}{2}, \frac{1}{2}, \frac{1}{2} \right)\, 
          \left( \frac{1}{2}, \frac{1}{2},   0,   0,   0,   0,-\frac{1}{2},-\frac{1}{2} \right) \;, \\
W_4 & = & W_6 ~=~ W_2\;,
\label{WevenId}
\end{eqnarray}
\end{subequations}
corresponding to the six torus directions $e_{\alpha}$.
These shifts and Wilson lines satisfy the  modular invariance conditions
\begin{equation}\label{eq:ModInv}
2 \left[ \left(k_1\,V_1+k_2\,V_2+n_\alpha\,W_\alpha\right)^2 - \left(k_1\,v_1+k_2\,v_2\right)^2\right] = 0 \mod 2
 \quad \forall~k_i,n_\alpha\in\{0,1\}\;,
\end{equation}
and, furthermore, fulfill the consistency requirements of
reference~\cite{Ploger:2007iq}. 
Equation~\eqref{eq:ModInv} is obtained by noticing that
the theta-functions inside the corresponding partition function are periodic
under the change of the modular parameter ($\rho \rightarrow \rho + 2$ for an
order two element) up to some phase factors that needs to be cancelled.

The states in the spectrum originate from different sectors: the untwisted
sectors $U_i$ (with $i=1,2,3$ corresponding to the $i^\mathrm{th}$ plane,
spanned by $e_{2i-1}$ and $e_{2i}$) and the twisted sectors $T_{(k,\ell)}$
(corresponding to the orbifold twist $k\, v_1 + \ell\, v_2$). In total, the
spectrum contains $6 \times\crep{10} + 15 \times\rep{5} + 9 \times\crep{5}$ of
$\SU5$, 52 non-Abelian singlets and some representations with respect to a
hidden sector gauge group $\SU4^2$. Three of the nine vector-like pairs of
$\rep{5}/\crep{5}$-plets are part of the untwisted sectors $U_i,~ i=1,2,3$,
originating from the 10D bulk; the remainder of the \SU5-charged spectrum
resides in the various twisted sectors. In particular, the six generations of
\SU5 are all twisted states.

\subsection{Modding out a freely acting $\boldsymbol{\Z2}$ involution}
\label{sec:freelyacting}

Next, we divide out the $\Z2$ symmetry corresponding to
\begin{equation}
\label{eqn:tau}
 \tau ~=~ \frac{1}{2}\left(e_2 + e_4 + e_6\right)
\end{equation}
with a gauge embedding denoted by $W$. Since $\tau$ acts freely, i.e.\ it does
not produce fixed points, we refer to $W$ as freely acting Wilson line. This is a
slight abuse of terminology, since (field-theoretic) Wilson lines are always
non-local. However, in the context of orbifold model building discrete Wilson
lines usually denote the differences between local shifts, i.e.\ they are Wilson
lines on the underlying torus but not on the orbifold (see e.g.\
\cite{GrootNibbelink:2003rc,Buchmuller:2006ik}). By contrast, $W$ is a Wilson
line also on the orbifold.

The strict identification of $W_2$, $W_4$ and $W_6$ in equation~\eqref{WevenId}
allows us to mod out $\tau$. Further, from its definition \eqref{eqn:tau} it
follows that $W$ is an element of order four,
\begin{equation}
 W~ =~ \frac{1}{2}\left( W_2 + W_4 + W_6 \right) ~=~ \frac{3}{2} W_2\;,
\end{equation}
as $W_2$ is of order two. 

Modular invariance of the resulting partition function for this order four
element amends the conditions \eqref{eq:ModInv} by 
\begin{equation}
 4\,\left(n_\alpha\,W_\alpha+n_0\,W\right)^2  ~=~0\mod 2  \quad\forall~n_0,n_\alpha\in\{0,1\}\;.
 \label{ModInvZ2free}
\end{equation}
In particular, we have chosen the Wilson line $W_2$ in equation~\eqref{eqn:W2}
such that $W$ satisfies all the conditions \eqref{ModInvZ2free} and breaks the
\SU5 GUT group down to $G_\mathrm{SM}$. 

By contrast, the Wilson line associated with an involution employed on smooth CY
are taken to be perpendicular to the gauge
bundle~\cite{Bouchard:2005ag,Braun:2005nv}. This is not the case in our
construction; precisely for that reason our Wilson line $W$ is an order 4
element instead of order 2.

\subsection{Massless spectrum}
\label{sec:masslessspectrum}

After modding out $\tau$, the 4D gauge group is  $G_\mathrm{SM}$ times eight
$\U1$ factors and a non-Abelian hidden sector $\SU3 \times \SU2 \times\SU2$. One
combination of the $\U1$ factors with generator $\mathsf{t}_\mathrm{anom}$
denotes the anomalous $\U1$. Furthermore, the standard hypercharge generator
$\mathsf{t}_Y$ from $\SU5$ can be identified and turns out to be orthogonal to
the anomalous direction,
\begin{subequations}
\begin{eqnarray}
\mathsf{t}_\mathrm{anom} & = & \left(-2,-1, 2, 1, 1, 1, 1, 1\right)\,
                             \left( -\frac{1}{2},-\frac{1}{2}, 0, 0, 0, 0,-\frac{1}{2},-\frac{1}{2}\right)\;, \\
\mathsf{t}_Y           & = & \left( 0,0,0,\frac{1}{2}, \frac{1}{2},-\frac{1}{3},-\frac{1}{3},-\frac{1}{3},\right)\,
                             \left( 0, 0, 0, 0, 0, 0, 0, 0\right)\;.
\end{eqnarray}
\end{subequations}
The model has a local \SU5 GUT structure (for the discussion of the concept of
local GUTs see~\cite{Buchmuller:2005sh} and cf.\ the related earlier discussion
in~\cite{Kobayashi:2004ya,Forste:2004ie}).

\begin{table}[t!]
\begin{center}
\begin{tabular}{|r|c|l|c|r|c|l|}
\hhline{|---|~|---|}
\# & representation & label & & \# & representation & label\\
\hhline{|---|~|---|}
$3$ & $(\crep{3}, \rep{2}; \rep{1}, \rep{1}, \rep{1})_{( \frac{1}{6}, \frac{1}{3})}$ & $q$ 
&& 
$3$ & $( \rep{3}, \rep{1}; \rep{1}, \rep{1}, \rep{1})_{(-\frac{2}{3},-\frac{1}{3})}$ & $\overline{u}$   \\
$3$ & $( \rep{3}, \rep{1}; \rep{1}, \rep{1}, \rep{1})_{( \frac{1}{3},-\frac{1}{3})}$ & $\overline{d}$ 
&&  
$3$ & $( \rep{1}, \rep{2}; \rep{1}, \rep{1}, \rep{1})_{(-\frac{1}{2}, -1)}$ &  $\ell$   \\
$3$ & $( \rep{1}, \rep{1}; \rep{1}, \rep{1}, \rep{1})_{( 1, 1)}$ &  $\overline{e}$ 
&&  
$33$ & $( \rep{1}, \rep{1}; \rep{1}, \rep{1}, \rep{1})_{(0,a)}$ & $s$   \\
$4$ & $( \rep{1}, \rep{2}; \rep{1}, \rep{1}, \rep{1})_{(-\frac{1}{2}, 0)}$ & $h$ 
&& 
$4$ & $( \rep{1}, \rep{2}; \rep{1}, \rep{1}, \rep{1})_{( \frac{1}{2}, 0)}$ & $\overline{h}$    \\
$5$ & $( \rep{3}, \rep{1}; \rep{1}, \rep{1}, \rep{1})_{( \frac{1}{3}, \frac{2}{3})}$ & $\overline{\delta}$ 
&& 
$5$ & $(\crep{3}, \rep{1};\rep{1}, \rep{1}, \rep{1})_{(-\frac{1}{3},-\frac{2}{3})}$ & $\delta$ \\
$5$ & $( \rep{1}, \rep{1}; \rep{3}, \rep{1}, \rep{1})_{(0,b)}$            &  $x$ 
&&  
$5$ & $( \rep{1}, \rep{1};\crep{3}, \rep{1}, \rep{1})_{(0,-b)}$            & $\overline{x}$ \\
$6$ & $( \rep{1}, \rep{1}; \rep{1}, \rep{1}, \rep{2})_{(0, 0)}$            &  $y$ 
&&  $6$ & $( \rep{1}, \rep{1}; \rep{1}, \rep{2}, \rep{1})_{(0, 0)}$            & $z$    \\
\hhline{|---|~|---|}
\end{tabular}
\end{center}
\caption{Spectrum at the orbifold point. We show the representations w.r.t.\
$G_\mathrm{SM}\times\U1_{B-L}\times[\SU3\times\SU2\times\SU2]_\mathrm{hid}$ and
their multiplicities (\#) and labels. The $[\dots]_\mathrm{hid}$ groups stem
from the second \E8, and $a \in\{0, \pm 1, \pm 2, \pm 3\}$ and $b \in
\{-4/3, -1/3, 5/3\}$. The $B-L$ generator is given in
equation~\eqref{B-L}.}
\label{tab:spectrum}
\end{table}

Dividing out the freely acting symmetry $\tau$ reduces the number of fixed
points from 48 to 24 and breaks the symmetry from \SU5 to $G_\mathrm{SM}$.  It
further splits the untwisted $\boldsymbol{5}$- and
$\overline{\boldsymbol{5}}$-plets in the $(e_1,e_2)$-plane to a pair of Higgs
candidates, denoted by $h_1$ and $\overline{h}_1$, removing the triplets. In
the other two planes it removes the doublets, leaving two pairs of
triplets/anti-triplets $\delta_i$ / $\overline{\delta}_i$ ($i =1,2$) massless. A
compact summary of the spectrum is given in table~\ref{tab:spectrum}; more
complete details have been listed in table~\ref{tab:fullspectrum} in
appendix~\ref{sec:FullSpectrum}.

To understand the family structure note that, due to the absence of the Wilson
line in the $e_1$ direction ($W_1 = 0$), states in the $T_{(0,1)}$ and
$T_{(1,1)}$ sectors form doublets under a discrete group $D_4$, which is
unaffected by modding out the freely acting symmetry $\tau$. As two families
reside in the $T_{(1,1)}$ sector, the two light families transform as a doublet
under this $D_4$ flavor symmetry. The third family comes from $T_{(1,0)}$ sector
and hence is a $D_4$ singlet. Such a $D_4$ symmetry is known to be
phenomenologically attractive as it can ameliorate supersymmetric flavor
problems \cite{Ko:2007dz}. In this respect the structure of the model is very
similar to the \Z6-II models discussed
in~\cite{Buchmuller:2006ik,Lebedev:2006tr,Lebedev:2007hv}.

\subsection{Gauge coupling unification}

As explained in subsection~\ref{sec:freelyacting}, the GUT symmetry breaking is
accomplished by the action of a freely acting symmetry $\tau$, leading to a
completely non-local breaking. This mechanism was introduced originally in the
context of smooth manifold compactifications~\cite{Witten:1985xc}. Later it was
considered as an alternative to the standard (localized) breaking in orbifold
constructions~\cite{Hebecker:2003we}. 

In order to discuss the virtues of non-local breaking, let us briefly recall the
usual obstructions in embedding the beautiful picture of MSSM gauge coupling
unification in the heterotic string. There are three main issues:
\begin{enumerate}
 \item huge, ``power-like'' threshold corrections around the string scale; 
 \item the MSSM unification scale, $M_\mathrm{GUT}=\text{few}\cdot10^{16}\,$GeV,
  is by an order 10 factor below the heterotic string scale;
 \item the appearance of split multiplets at the high scale generically leads to
  logarithmic thresholds.
\end{enumerate}
The first problem is absent in the scheme of `local grand unification' as the
bulk gauge group in extra dimensions contains $G_\mathrm{SM}$ such that
power-like corrections are universal. The second issue may be overcome by
considering anisotropic compactifications
\cite{Witten:1996mzF3,Hebecker:2004ce}. As described in detail in
\cite{Hebecker:2004ce}, by using a discrete (rather than continuous) Wilson line
associated with the involution to break the GUT symmetry, the breaking scale is
related to the length of the corresponding Wilson line cycle. This length can be
of order of $M_\mathrm{GUT}^{-1}$ with the volume of compact space being so
small that a description in terms of the perturbative heterotic string is still
justifiable. This mechanism also ameliorates the third problem. In fact, the
remaining logarithmic corrections may even mitigate the discrepancy between
string and GUT scales \cite{Ross:2004mi}. In this respect our model is
``cleaner'' than the MSSMs based on \Z{6}-II, where various logarithmic
corrections to gauge unification from localized states and vector-like exotics
coming in incomplete GUT multiplets are expected (cf.\ the discussion
in~\cite{Dundee:2008ts}). Hence, the mechanism of non-local GUT breaking
provides us with one of the most compelling realizations of precision gauge
coupling unification. The implications of precision gauge unification for
the MSSM superpartner spectrum have been
discussed very recently in \cite{Raby:2009sf}.

\section{Semi-realistic VEV configuration}
\label{sec:phenoVEVs}

In order to obtain the MSSM, the unwanted \U1 gauge group factors have to be
broken. This can be accomplished
by switching on VEVs of standard model singlet fields consistently with
vanishing $F$- and $D$-terms (cf.\ the discussion
in~\cite{Buchmuller:2006ik,Lebedev:2007hv}). In addition, these VEVs give rise
to effective Yukawa couplings for quarks and leptons, they serve as effective
mass terms decoupling the exotics
and may generate dangerous proton decay operators. In order to avoid the latter ones
(at least of dimension four), we identify vacuum configurations with a
matter parity, using methods described
in~\cite{Petersen:2009ip}. Like in the heterotic benchmark model
in~\cite{Lebedev:2007hv}, this matter parity emerges as a \Z2 subgroup of a
$\U1_{B-L}$ gauge symmetry generated by
\begin{equation}
 \mathsf{t}_{B-L}~=~
 \left(-\frac{1}{2}, \frac{1}{2},-\frac{1}{2}, \frac{1}{2}, \frac{1}{2},-\frac{1}{6},-\frac{1}{6},-\frac{1}{6} \right)\,
 \left(\frac{13}{6},-\frac{7}{6}, \frac{5}{6}, \frac{5}{6}, \frac{5}{6}, \frac{5}{6},-\frac{1}{2},-\frac{1}{2} \right)\;,
 \label{B-L} 
\end{equation}
and is given by $\mathrm{e}^{2\pi\I \frac{3}{2}q_{B-L}} = \pm 1$. This matter
parity will be referred to as $\Z2^R$. The configurations with preserved $\Z2^R$
are such that we are left with the exact MSSM gauge group, three chiral
generations, no $R$ parity violating couplings, and are able to discriminate
between lepton and Higgs doublets as well as between SM fields and exotics (see
table~\ref{tab:spectrum}).

Let us now discuss a configuration in which all
$G_\mathrm{SM}\times\Z2^R$ singlets $\phi^{(i)}$ are assumed to attain
VEVs
\begin{eqnarray}
\label{eqn:vevconfig}
\{\phi^{(i)}\}
& = & \{s_1, s_2, s_3, s_4, s_5, s_7, s_8, s_9, s_{10}, s_{15}, s_{16}, s_{17}, s_{18}, s_{19}, s_{20}, s_{21}, \nonumber \\
&   &   s_{22},s_{23},s_{25},s_{26},s_{27},s_{28},s_{30},s_{31},s_{32},s_{33},x_3,x_4,x_5,\overline{x}_2,\overline{x}_4,\overline{x}_5,\nonumber \\
&   &   y_1,y_2,y_3,y_4,y_5,y_6,z_1,z_2,z_3,z_4,z_5,z_6 \}\;.
\end{eqnarray}
Configurations in which all these 44 fields are non-trivial lead to 44 $F$-term
equations for 44 fields, which in general have solutions (cf.\ the corresponding
discussion in \cite{Buchmuller:2006ik}). Furthermore, we have explicitly
verified that these fields enter gauge invariant monomials, such as to ensure
vanishing $D$-terms including a cancellation of the FI term of the anomalous
\U1. Due to our ignorance of the coefficients of couplings we were not able to
prove that the simultaneous solutions to $F=D=0$ occur for  small singlet
expectation values; in the following we make this assumption.

Assigning VEVs to all the $\phi^{(i)}$ fields breaks all extra \U1 factors and
leads to effective mass terms for the non-chiral remnants w.r.t.\ the symmetry
$G_\mathrm{SM}\times\Z2^R$. This can be seen in detail by generating
all couplings allowed by the relevant string selection rules, and compute the
corresponding mass matrices $\mathcal{M}_{ij}$. As the freely acting symmetry
slightly modifies the usual selection rules, we specify them explicitly in
appendix~\ref{app:SelectionRules}. For the exotic
$\delta_i$-$\overline{\delta}_j$ pairs we obtain the structure
\begin{equation}\label{eq:Mdelta}
\mathcal{M}^\delta_{ij} ~\sim~ \left(
\begin{array}{ccccc}
\phi^3 & s_1    &  \phi^3 & \phi^3 & \phi^3    \\
s_2    & \phi^3 &  \phi^5 & s_{16} & s_{20} \\
\phi^5 & \phi^3 &  \phi^5 & s_{26} & s_{31} \\
s_{28} & \phi^3 &  s_{19} & s_{10} & \phi^3    \\
s_{33} & \phi^3 &  s_{23} & \phi^3 & s_{10} \\
\end{array}
\right)\;, 
\end{equation}
where here and in the following $\phi^n$ denotes a sum of known monomials in the
VEVs of the fields of \eqref{eqn:vevconfig} with $n$ being its lowest degree.
Obviously, due to the $\Z2^R$ matter parity, there is no mixing
between $\overline{d}$ quarks and the quark-like exotics $\overline{\delta}$. 
Switching on the VEVs of the untwisted states $s_1$ and  $s_2$ is sufficient to
decouple the untwisted triplets $\delta_i$, $\overline{\delta}_i$  $i=1,2$.
Even more, as can be seen from \eqref{eq:Mdelta}, all triplets decouple at linear order
in the $\phi^{(i)}$ fields. Similar features have been reported in the
context of free fermionic model building (see e.g.\ \cite{Cleaver:1998saa}).

There are four Higgs pair candidates with mass matrix, defined by the
superpotential terms $h_i\,\mathcal{M}^h_{ij}\, \overline{h}_j$,
\begin{equation}
\mathcal{M}^h_{ij} \sim \left(
\begin{array}{cccc}
\phi^3 & s_3    & \phi^3 & \phi^3 \\
s_{15} & \phi^5 & s_{19} & s_{23} \\
\phi^3 & s_{26} & s_{10} & \phi^3 \\
\phi^3 & s_{31} & \phi^3 & s_{10} \\
\end{array}
\right)
\end{equation}
of maximal rank. Thus, this VEV configuration suffers under the stringy version
of the $\mu$ problem. The Yukawa couplings of the quarks 
($q_i\,\mathcal{M}^{\overline{u}}_{ij}\, \overline{u}_j$ and 
$q_i\,\mathcal{M}^{\overline{d}}_{ij}\, \overline{d}_j$) and of  the charged
leptons ($\ell_i\,\mathcal{M}^{\overline{e}}_{ij}\, \overline{e}_j$) are of the
form
\begin{equation}
\label{eqn:yuk}
\mathcal{M}^{\overline{u}} \sim \left(
\begin{array}{ccc}
\overline{h} \phi^4 & \overline{h} \phi^4 & 0 \\
\overline{h} \phi^4 & \overline{h} \phi^4 & 0 \\
0                   & 0                   & \overline{h}_1\\
\end{array}
\right) \quad\text{and}\quad 
\mathcal{M}^{\overline{d}} \sim \mathcal{M}^{\overline{e}} \sim \left(
\begin{array}{ccc}
0   & 0   & h_3 \\
0   & 0   & h_4 \\
h_3 & h_4 & 0   \\
\end{array}
\right)\;.
\end{equation}
Generically, they depend on the VEVs of all four Higgs-pairs and,  due to
$\Z2^R$, there is no mixing between the lepton-doublets $\ell$ and the
Higgses $h$. The top-quark couples to $\overline{h}_1$ already at order
$\phi^0$, hence this coupling is not suppressed compared to the first and second
generations. Moreover, as the Higgs $\overline{h}_1$ is part of the untwisted
sector, i.e.\ an internal part of the 10D gauge field, it couples with a
strength proportional to the gauge coupling, realizing a gauge-top
unification~\cite{Hosteins:2009xk}.

In general, couplings between localized states exhibit \SU5 relations, as they
are not subject to the non-local symmetry breakdown due to $W$.  Furthermore, as
a consequence of our choice of $\U1_{B-L}$, the three generations of quarks and
leptons originate only from the twisted sectors, hence their couplings originate
from \SU5. This explains why the charged lepton mass matrix
$\mathcal{M}^{\overline{e}}$ and the $d$-type mass matrix
$\mathcal{M}^{\overline{d}}$ are identical in equation~\eqref{eqn:yuk}, a
feature that is actually only desirable for the third generation.

\section{Interpretation as a complete blow-up}
\label{sec:blowup}

The second objective of our work is to show that our orbifold model may be
related to a smooth Calabi-Yau compactification. The $\Z2\times\Z2$ orbifold has
48 $\Z2$ fixed tori that constitute codimension four singularities, which are
identified pairwise by the freely acting symmetry. To obtain a smooth space all
these singularities have to be polished out. Below we will explain why the
configuration discussed in the previous section defines a complete blow-up
within the effective 4D theory.

From the perspective of the orbifold model smoothing out singularities
corresponds to non-vanishing VEVs of twisted states. To smooth out all
singularities at least one twisted state per fixed torus needs to acquire a VEV. 
Such a VEV can either lead to a blow-up or to a deformation of these
singularities~\cite{Vafa:1994rv}. In the former case the cycle hidden inside the
singularity, called the exceptional divisor, acquires a finite volume w.r.t.\
the K\"ahler form of the geometry. When the singularity is deformed, i.e.\ the
complex structure is modified, its volume remains zero, and is in this sense
still singular. As all twisted states of the $\Z{2}\times\Z{2}$ orbifold are six
dimensional, they form hyper multiplets of $\mathcal{N}=1$ in 6D. Which of the
two chiral multiplets within these hyper multiplets takes a VEV decides whether
one has a blow-up or a deformation.

The blow-up described within the effective 4D theory takes into account only
those twisted states as blow-up-modes that are massless in 4D. Due to the
presence of Wilson lines it may happen that some orbifold singularities do not
provide 4D massless states. In fact, a quick glance over
table~\ref{tab:fullspectrum} in appendix~\ref{sec:FullSpectrum} reveals that
three fixed tori of the orbifold model defined in section~\ref{sec:model} do not
support 4D zero modes. So they might remain singular in a complete blow-up
within the effective 4D theory. On the other hand, each fixed torus supports 6D
massless twisted states, which may develop non-trivial profiles over the
internal tori that might remove the singularities. However, a detailed discussion of
these issues is beyond the scope of the present letter. 

In general, VEV configurations correspond to complicated gauge bundles on the
Calabi-Yau space that need to fulfill the integrated Bianchi identities 
\begin{equation}
\int_S ( \tr \mathcal{R}^2 - \tr \mathcal{F}^2) ~=~ 0
\end{equation}
for all four-cycles $S$. These consistency equations can be viewed as the smooth
analog of the modular invariance conditions, equation~\eqref{eq:ModInv}, for the
shifts $V_1, V_2$ and the Wilson lines $W_\alpha$. However, there seems to be no
condition(s) on the Wilson line $W$ associated with the involution $\tau$ for
our blow-up or for other smooth CY constructions. By contrast, on the orbifold
we encounter the additional requirements~\eqref{ModInvZ2free}. Their derivation
necessarily involves winding modes. The fact that in the supergravity
approximation they are usually ignored, might explain why the modular invariance
conditions of the freely acting Wilson line $W$ do not have a smooth
counterpart. Nevertheless, such conditions might be essential to ensure that a
given smooth CY compactification of supergravity has a full string lift.

The VEV configuration~\eqref{eqn:vevconfig} has been chosen such that  the
standard model group and matter parity (in particular also the hypercharge)
remain unbroken. Moreover, according to table~\ref{tab:fullspectrum} in
appendix~\ref{sec:FullSpectrum} all non-empty fixed tori support twisted states
to the 4D theory with VEVs switched on. Therefore, this corresponds to a
complete blow-up in the effective 4D theory. Hence, it shows that the
obstructions to a full blow-up encountered in the \Z{6}-II
mini-landscape models can be overcome in settings with non-local GUT breaking.

To summarize, we have shown that the $\Z2\times\Z2$ orbifold model with a freely
acting $\Z2$ involution allows for VEV configurations where 4D zero modes
originating from all non-empty fixed tori are switched on without breaking the
standard model gauge group. The construction and interpretation of such
configuration from the point of view of smooth compactifications will be
discussed elsewhere~\cite{Blaszczyk:2009pr}.

\section{Conclusions}
\label{sec:conclusions}

We have presented a $\Z2\times\Z2$ orbifold compactification of the 
heterotic string exhibiting the exact chiral MSSM spectrum and gauge group as
well as matter parity. 
The starting point of this model is the $\Z2\times\Z2$ orbifold with \SU5 gauge
group. The \SU5 GUT symmetry is non-locally broken by modding out a
freely acting symmetry. This ensures that there is no flux in hypercharge
direction such that there is no obstruction to a complete blow-up. Further,
Wilson line breaking is known to avoid large thresholds to the gauge coupling
such that our construction complies with the beautiful picture of MSSM gauge
coupling unification.

Accompanying an involution of the geometry
with a Wilson line has been considered previously in smooth compactifications
leading to the MSSM~\cite{Bouchard:2005ag,Braun:2005nv}. 
However, in our approach we encounter novel modular invariance conditions on
this freely acting Wilson line that seem to have no analog in smooth CY
compactifications in the supergravity approximation. 
This might suggest that some of the smooth CYs with involutions dressed with
Wilson lines may exist only as supergravity models, but do not have a lift to
consistent string theory constructions.

The model has the chiral MSSM spectrum and many other phenomenologically
appealing features, like non-trivial Yukawa couplings and admits vacua with
matter parity. On the other hand, we cannot claim that the configuration
presented here is fully realistic. In more detail,
due to the presence of a $D$-term for an anomalous \U1 some states need to
acquire VEVs. We identified and discussed a specific VEV configuration with an
exact matter parity (hence proton decay is avoided at the dimension four level),
where all unwanted \U1 gauge group factors are broken, and all exotics decouple.
Unfortunately, also the Higgs fields generically attain large masses. This
unpleasant feature is shared with smooth CY MSSM
models~\cite{Bouchard:2005ag,Anderson:2009mh}, where generically the $\mu$-term
is of the order of the fundamental
scale~\cite{Bouchard:2006dn,Anderson:2009mh} (while in orbifold models there are
symmetries that allow us to relate the size of $\mu$ to the scale of
supersymmetry breakdown~\cite{Lebedev:2007hv,Buchmuller:2008uq,Kappl:2008ie}).  Our model also avoids the problem
of an additional $\U1_{B-L}$ symmetry that cannot be broken without breaking
supersymmetry. 
In summary, we have presented an explicit orbifold compactification
satisfying all stringy consistency conditions. We identified vacua which
correspond to resolutions of the orbifold fixed points, have properties very
similar to those of the most promising smooth heterotic compactifications
known so far, and are, in addition, endowed with an exact matter parity.

\subsubsection*{Outlook}

The main achievement in this letter was to show how to construct a 
concrete orbifold compactification of the heterotic string in which the breaking 
$\SU5\to G_\mathrm{SM}$ is non-local, i.e.\ due to a Wilson line.
We have argued that this may allow us to obtain a potentially realistic
model with an explicit orbifold limit and a clear interpretation in terms of
smooth geometry.
Our analysis is incomplete in three main respects. First, the phenomenological
viability of the model has to be studied in more detail. The configuration
discussed in this letter suffers from the problem that the Higgses generically
get ultra-heavy. Possible solutions to the $\mu$-problem will be discussed in a
forthcoming publication~\cite{Kappl:2009pr2}.

Secondly, the configuration with VEVs discussed in this work seems to indicate
that a complete blow-up within the effective 4D theory is possible. However to
really show that this configuration corresponds to a smooth compactification,
one has to construct the gauge bundle on the resolution of the compact orbifold
explicitly and check that it fulfills all Bianchi identities for consistency.
Work in this direction is in progress~\cite{Blaszczyk:2009pr}.

Finally, we have seen that some twisted sectors are empty in 4D. This seems to
indicate that the corresponding orbifold singularity remain unresolved.
Therefore, they correspond to partial (rather than full) blow-ups of the
geometry. A geometric interpretation of such settings still needs to be
obtained.

\subsubsection*{Acknowledgments}

We would like to thank G.~Curio, H.P.~Nilles and G.~Ross for useful 
discussions. This research was supported by the DFG cluster of excellence Origin
and Structure of the Universe, the European Union 6th framework program
\mbox{MRTN-CT-2006-035863} ``UniverseNet'', LMUExcellent and the
\mbox{SFB-Transregios} 27 ``Neutrinos and Beyond'' and 33 ``The Dark Universe''
by Deutsche Forschungsgemeinschaft (DFG). One of us (M.R.) would like to thank
the Aspen Center for Physics, where part of this work was done, for hospitality
and support. The work of M.T.\ was supported by the EC under the Marie Curie
Host Fellowship for the Transfer of Knowledge, contract n.\ MTKD-CT-2005-029466.

\appendix

\section{Full spectrum}
\label{sec:FullSpectrum} 

In table~\ref{tab:fullspectrum}, we present in detail the spectrum of the model
discussed in section~\ref{sec:model}. We decompose the states in untwisted
($U_i$ with $i=1,2,3$ corresponding to the $i^\mathrm{th}$ plane) and twisted
($T_{(k,\ell)}^{(n_1,n_2,\dots, n_6)}$) sectors, where $(k,\ell)$ and
$(n_1,n_2,\dots, n_6)$ indicate the corresponding ``constructing elements'' (of
twisted strings with boundary conditions $X(\tau,\sigma+2\pi) =
\theta^k\,\omega^\ell\, X(\tau,\sigma) + n_{\alpha}\, e_{\alpha}$ with $\theta$
and $\omega$ denoting the rotations corresponding to $v_1$ and $v_2$,
respectively). As all twisted states live on two-tori in six dimensions, we
indicate the directions $n_i$ where these tori lie by $*$. The states that
acquire a VEV in the configuration discussed in section~\ref{sec:phenoVEVs} are
indicated with angular brackets $\langle \ \rangle$.  
{\scriptsize
\begin{longtable}{|l|l|c|ccccccccc|c|}
\caption{Spectrum of the model at the orbifold point. } 
\label{tab:fullspectrum}
\\
\hline
sector & irrep &$R_1, R_2, R_3$ &
$\!q_\text{anom}\!$ & $\!q_{1}\!$ & $\!q_{2}\!$ & $\!q_{3}\!$ & $\!q_{4}\!$ 
& $\!q_{5}\!$ & $\!q_{6}\!$ & $\!q_{7}\!$ & $\!q_{8}\!$ & label(s) \\
\hline
\hline
\endfirsthead
\hline
sector & irrep &$R_1, R_2, R_3$ &
$\!q_\text{anom}\!$ & $\!q_{1}\!$ & $\!q_{2}\!$ & $\!q_{3}\!$ & $\!q_{4}\!$ 
& $\!q_{5}\!$ & $\!q_{6}\!$ & $\!q_{7}\!$ & $\!q_{8}\!$ & label(s) \\
\hline
\hline
\endhead
\hline
\multicolumn{13}{|r|}{{\it continued ...}}\\
\hline
\endfoot
\hline
\endlastfoot
$U_1$  & $(\rep{1}, \rep{1}, \rep{1}, \rep{1}, \rep{1})$ & $-1, 0, 0$ & $ 2$ & $ 0$ & $ 2$ & $ 4$ & $ 20$ & $ 7$ & $ -37$ & $ 0$ & $ 0$ &  $\langle s_{1}\rangle$\\
 & $(\rep{1}, \rep{2}, \rep{1}, \rep{1}, \rep{1})$ & $-1, 0, 0$ & $ 2$ & $-\tfrac{1}{2}$ & $ 2$ & $ 4$ & $ -9$ & $ 6$ & $ 6$ & $ -4$ & $ 0$ &  $h_{1}$\\
 & $(\rep{1}, \rep{2}, \rep{1}, \rep{1}, \rep{1})$ & $-1, 0, 0$ & $ -2$ & $\tfrac{1}{2}$ & $ -2$ & $ -4$ & $ 9$ & $ -6$ & $ -6$ & $ 4$ & $ 0$ &  $\overline{h}_{1}$\\
 & $(\rep{1}, \rep{1}, \rep{1}, \rep{1}, \rep{1})$ & $-1, 0, 0$ & $ -2$ & $ 0$ & $ -2$ & $ -4$ & $ -20$ & $ -7$ & $ 37$ & $ 0$ & $ 0$ &  $\langle s_{2} \rangle$ \\
$U_2$  & $(\crep{3}, \rep{1}, \rep{1}, \rep{1}, \rep{1})$ & $0, -1, 0$ & $ -1$ & $-\tfrac{1}{3}$ & $ -1$ & $ -2$ & $-\tfrac{107}{2}$ & $ -5$ & $ -27$ & $-\tfrac{1}{2}$ & $ 0$ &  $\delta_{2}$\\
 & $(\rep{3}, \rep{1}, \rep{1}, \rep{1}, \rep{1})$ & $0, -1, 0$ & $ 1$ & $\tfrac{1}{3}$ & $ 1$ & $ 2$ & $\tfrac{107}{2}$ & $ 5$ & $ 27$ & $\tfrac{1}{2}$ & $ 0$ &  $\overline{\delta}_{2}$\\
$U_3$  & $(\rep{3}, \rep{1}, \rep{1}, \rep{1}, \rep{1})$ & $0, 0, -1$ & $ 3$ & $\tfrac{1}{3}$ & $ 3$ & $ 6$ & $\tfrac{147}{2}$ & $ 12$ & $ -10$ & $\tfrac{1}{2}$ & $ 0$ &  $\overline{\delta}_{1}$\\
 & $(\crep{3}, \rep{1}, \rep{1}, \rep{1}, \rep{1})$ & $0, 0, -1$ & $ -3$ & $-\tfrac{1}{3}$ & $ -3$ & $ -6$ & $-\tfrac{147}{2}$ & $ -12$ & $ 10$ & $-\tfrac{1}{2}$ & $ 0$ &  $\delta_{1}$\\
\hline
\hline
$T_{(1, 0)}^{(*,*, 0, 0, 0, 0)}$  & $(\rep{1}, \rep{2}, \rep{1}, \rep{1}, \rep{1})$ & $0, -\tfrac{1}{2}, -\tfrac{1}{2}$ & $ 2$ & $\frac{1}{2}$ & $ 2$ & $ 4$ & $-\tfrac{47}{2}$ & $\frac{11}{2}$ & $\frac{11}{2}$ & $\frac{5}{2}$ & $ 0$ &  $\overline{h}_{2}$\\
 & $(\crep{3}, \rep{1}, \rep{1}, \rep{1}, \rep{1})$ & $0, -\tfrac{1}{2}, -\tfrac{1}{2}$ & $ 2$ & $-\tfrac{1}{3}$ & $ 2$ & $ 4$ & $-\tfrac{47}{2}$ & $\frac{11}{2}$ & $\frac{11}{2}$ & $\frac{5}{2}$ & $ 0$ &  $\delta_{3}$\\
 & $(\rep{1}, \rep{1}, \rep{1}, \rep{1}, \rep{1})$ & $0, -\tfrac{1}{2}, -\tfrac{1}{2}$ & $ -4$ & $ 0$ & $ -4$ & $ -8$ & $\frac{65}{2}$ & $-\tfrac{23}{2}$ & $-\tfrac{23}{2}$ & $\frac{3}{2}$ & $ 0$ &  $\langle s_{3} \rangle $\\
 & $(\crep{3}, \rep{2}, \rep{1}, \rep{1}, \rep{1})$ & $0, -\tfrac{1}{2}, -\tfrac{1}{2}$ & $ 1$ & $\frac{1}{6}$ & $ 1$ & $ 2$ & $-\tfrac{9}{2}$ & $ 3$ & $ 3$ & $ -2$ & $ 0$ &  $q_{3}$\\
 & $(\rep{3}, \rep{1}, \rep{1}, \rep{1}, \rep{1})$ & $0, -\tfrac{1}{2}, -\tfrac{1}{2}$ & $ 1$ & $-\tfrac{2}{3}$ & $ 1$ & $ 2$ & $-\tfrac{9}{2}$ & $ 3$ & $ 3$ & $ -2$ & $ 0$ &  $\overline{u}_{3}$\\
 & $(\rep{1}, \rep{1}, \rep{1}, \rep{1}, \rep{1})$ & $0, -\tfrac{1}{2}, -\tfrac{1}{2}$ & $ 1$ & $ 1$ & $ 1$ & $ 2$ & $-\tfrac{9}{2}$ & $ 3$ & $ 3$ & $ -2$ & $ 0$ &  $\overline{e}_{3}$\\
\hline
$T_{(1, 0)}^{(*,*, 0, 0, 0, 1)}$ & empty &&&&&&&&&&& \\
\hline
$T_{(1, 0)}^{(*,*, 0, 0, 1, 0)}$  & $(\rep{1}, \rep{1}, \crep{3}, \rep{1}, \rep{1})$ & $0, -\tfrac{1}{2}, -\tfrac{1}{2}$ & $ 1$ & $ 0$ & $ -9$ & $ 31$ & $ 10$ & $\frac{7}{2}$ & $-\tfrac{37}{2}$ & $ 0$ & $ 0$ &  $\overline{x}_{1}$\\
 & $(\rep{1}, \rep{1}, \rep{3}, \rep{1}, \rep{1})$ & $0, -\tfrac{1}{2}, -\tfrac{1}{2}$ & $ 1$ & $ 0$ & $ 11$ & $ -27$ & $ 10$ & $\frac{7}{2}$ & $-\tfrac{37}{2}$ & $ 0$ & $ 0$ &  $x_{1}$\\
 & $(\rep{1}, \rep{1}, \rep{1}, \rep{1}, \rep{1})$ & $0, -\tfrac{1}{2}, -\tfrac{1}{2}$ & $ 0$ & $ 0$ & $ -15$ & $ -30$ & $ -5$ & $-\tfrac{95}{2}$ & $\frac{37}{2}$ & $ 0$ & $ 0$ &  $\langle s_{4} \rangle $\\
 & $(\rep{1}, \rep{1}, \rep{1}, \rep{1}, \rep{1})$ & $0, -\tfrac{1}{2}, -\tfrac{1}{2}$ & $ -2$ & $ 0$ & $ 13$ & $ 26$ & $ -15$ & $\frac{81}{2}$ & $\frac{37}{2}$ & $ 0$ & $ 0$ &  $\langle s_{5} \rangle $\\
 \hline 
$T_{(1, 0)}^{(*,*, 0, 0, 1, 1)}$  & $(\rep{1}, \rep{1}, \rep{3}, \rep{1}, \rep{1})$ & $0, -\tfrac{1}{2}, -\tfrac{1}{2}$ & $ 1$ & $ 0$ & $-\tfrac{23}{2}$ & $\frac{3}{2}$ & $ 80$ & $-\tfrac{71}{4}$ & $\frac{17}{4}$ & $\frac{1}{4}$ & $ 0$ &  $x_{2}$\\
 & $(\rep{1}, \rep{1}, \rep{1}, \rep{1}, \rep{1})$ & $0, -\tfrac{1}{2}, -\tfrac{1}{2}$ & $ 1$ & $ 0$ & $\frac{17}{2}$ & $-\tfrac{113}{2}$ & $ 80$ & $-\tfrac{71}{4}$ & $\frac{17}{4}$ & $\frac{1}{4}$ & $ 0$ &  $s_{6}$\\
 & $(\rep{1}, \rep{1}, \rep{3}, \rep{1}, \rep{1})$ & $0, -\tfrac{1}{2}, -\tfrac{1}{2}$ & $ 0$ & $ 0$ & $-\tfrac{25}{2}$ & $-\tfrac{1}{2}$ & $ -75$ & $-\tfrac{105}{4}$ & $-\tfrac{17}{4}$ & $-\tfrac{1}{4}$ & $ 0$ &  $\langle x_{3} \rangle $\\
 & $(\rep{1}, \rep{1}, \rep{1}, \rep{1}, \rep{1})$ & $0, -\tfrac{1}{2}, -\tfrac{1}{2}$ & $ 0$ & $ 0$ & $\frac{15}{2}$ & $-\tfrac{117}{2}$ & $ -75$ & $-\tfrac{105}{4}$ & $-\tfrac{17}{4}$ & $-\tfrac{1}{4}$ & $ 0$ &  $\langle s_{7} \rangle $\\
 \hline 
$T_{(1, 0)}^{(*, *, 1, 0, 0, 0)}$  & $(\rep{1}, \rep{1}, \rep{1}, \rep{1}, \rep{1})$ & $0, -\tfrac{1}{2}, -\tfrac{1}{2}$ & $ 2$ & $ 0$ & $-\tfrac{11}{2}$ & $\frac{125}{2}$ & $\frac{45}{2}$ & $ -15$ & $ 7$ & $\frac{3}{2}$ & $ 0$ &  $\langle s_{8} \rangle $\\
 & $(\rep{1}, \rep{1}, \crep{3}, \rep{1}, \rep{1})$ & $0, -\tfrac{1}{2}, -\tfrac{1}{2}$ & $ 2$ & $ 0$ & $\frac{29}{2}$ & $\frac{9}{2}$ & $\frac{45}{2}$ & $ -15$ & $ 7$ & $\frac{3}{2}$ & $ 0$ &  $\langle\overline{x}_{2}\rangle$\\
 & $(\rep{1}, \rep{1}, \rep{1}, \rep{1}, \rep{2})$ & $0, -\tfrac{1}{2}, -\tfrac{1}{2}$ & $\frac{3}{2}$ & $ 0$ & $ 9$ & $ 18$ & $\frac{35}{2}$ & $ 29$ & $ 7$ & $\frac{3}{2}$ & $-\tfrac{1}{2}$ &  $\langle y_{1} \rangle $\\
 & $(\rep{1}, \rep{1}, \rep{1}, \rep{2}, \rep{1})$ & $0, -\tfrac{1}{2}, -\tfrac{1}{2}$ & $\frac{3}{2}$ & $ 0$ & $ 9$ & $ 18$ & $\frac{35}{2}$ & $ 29$ & $ 7$ & $\frac{3}{2}$ & $\frac{1}{2}$ &  $\langle z_{1} \rangle $\\
 \hline 
$T_{(1, 0)}^{(*, *, 1, 0, 0, 1)}$  & $(\rep{1}, \rep{1}, \rep{1}, \rep{1}, \rep{1})$ & $0, -\tfrac{1}{2}, -\tfrac{1}{2}$ & $ 2$ & $ 0$ & $ 17$ & $ 34$ & $-\tfrac{95}{2}$ & $\frac{25}{4}$ & $-\tfrac{63}{4}$ & $\frac{5}{4}$ & $ 0$ &  $\langle s_{9} \rangle $\\
 & $(\rep{1}, \rep{1}, \rep{1}, \rep{1}, \rep{1})$ & $0, -\tfrac{1}{2}, -\tfrac{1}{2}$ & $ 0$ & $ 0$ & $ -15$ & $ -30$ & $\frac{135}{2}$ & $\frac{3}{4}$ & $-\tfrac{85}{4}$ & $-\tfrac{5}{4}$ & $ 0$ &  $\langle s_{10} \rangle $\\
 & $(\rep{1}, \rep{2}, \rep{1}, \rep{1}, \rep{1})$ & $0, -\tfrac{1}{2}, -\tfrac{1}{2}$ & $ 0$ & $-\tfrac{1}{2}$ & $ -15$ & $ -30$ & $ -34$ & $-\tfrac{11}{4}$ & $-\tfrac{11}{4}$ & $-\tfrac{5}{4}$ & $ 0$ &  $\ell_{3}$\\
 & $(\rep{1}, \rep{1}, \rep{1}, \rep{1}, \rep{1})$ & $0, -\tfrac{1}{2}, -\tfrac{1}{2}$ & $ 2$ & $ 0$ & $ 17$ & $ 34$ & $ 25$ & $\frac{35}{4}$ & $\frac{35}{4}$ & $-\tfrac{11}{4}$ & $ 0$ &  $s_{11}$\\
 & $(\rep{3}, \rep{1}, \rep{1}, \rep{1}, \rep{1})$ & $0, -\tfrac{1}{2}, -\tfrac{1}{2}$ & $ 0$ & $\frac{1}{3}$ & $ -15$ & $ -30$ & $ -34$ & $-\tfrac{11}{4}$ & $-\tfrac{11}{4}$ & $-\tfrac{5}{4}$ & $ 0$ &  $\overline{d}_{3}$\\
 \hline 
$T_{(1, 0)}^{(*, *, 1, 0, 1, 0)}$  & $(\rep{1}, \rep{1}, \rep{1}, \rep{1}, \rep{1})$ & $0, -\tfrac{1}{2}, -\tfrac{1}{2}$ & $ 2$ & $ 0$ & $-\tfrac{41}{2}$ & $\frac{65}{2}$ & $\frac{35}{2}$ & $ 29$ & $ 7$ & $\frac{3}{2}$ & $ 0$ &  $s_{12}$\\
 & $(\rep{1}, \rep{1}, \crep{3}, \rep{1}, \rep{1})$ & $0, -\tfrac{1}{2}, -\tfrac{1}{2}$ & $ 2$ & $ 0$ & $-\tfrac{1}{2}$ & $-\tfrac{51}{2}$ & $\frac{35}{2}$ & $ 29$ & $ 7$ & $\frac{3}{2}$ & $ 0$ &  $\overline{x}_{3}$\\
 & $(\rep{1}, \rep{1}, \rep{1}, \rep{2}, \rep{1})$ & $0, -\tfrac{1}{2}, -\tfrac{1}{2}$ & $\frac{3}{2}$ & $ 0$ & $ 9$ & $ 18$ & $\frac{35}{2}$ & $ 29$ & $ 7$ & $\frac{3}{2}$ & $-\tfrac{1}{2}$ &  $\langle z_{2} \rangle $\\
 & $(\rep{1}, \rep{1}, \rep{1}, \rep{1}, \rep{2})$ & $0, -\tfrac{1}{2}, -\tfrac{1}{2}$ & $\frac{3}{2}$ & $ 0$ & $ 9$ & $ 18$ & $\frac{35}{2}$ & $ 29$ & $ 7$ & $\frac{3}{2}$ & $\frac{1}{2}$ &  $\langle y_{2} \rangle $\\
 \hline 
$T_{(1, 0)}^{(*, *, 1, 0, 1, 1)}$  & $(\rep{1}, \rep{1}, \rep{1}, \rep{1}, \rep{1})$ & $0, -\tfrac{1}{2}, -\tfrac{1}{2}$ & $ 2$ & $ 0$ & $ 2$ & $ 4$ & $-\tfrac{105}{2}$ & $\frac{201}{4}$ & $-\tfrac{63}{4}$ & $\frac{5}{4}$ & $ 0$ &  $s_{13}$\\
 & $(\rep{1}, \rep{1}, \rep{1}, \rep{1}, \rep{1})$ & $0, -\tfrac{1}{2}, -\tfrac{1}{2}$ & $ 0$ & $ 0$ & $ 0$ & $ 0$ & $\frac{145}{2}$ & $-\tfrac{173}{4}$ & $-\tfrac{85}{4}$ & $-\tfrac{5}{4}$ & $ 0$ &  $s_{14}$\\
 & $(\rep{1}, \rep{2}, \rep{1}, \rep{1}, \rep{1})$ & $0, -\tfrac{1}{2}, -\tfrac{1}{2}$ & $ 0$ & $-\tfrac{1}{2}$ & $ 0$ & $ 0$ & $ -29$ & $-\tfrac{187}{4}$ & $-\tfrac{11}{4}$ & $-\tfrac{5}{4}$ & $ 0$ &  $h_{2}$\\
 & $(\rep{1}, \rep{1}, \rep{1}, \rep{1}, \rep{1})$ & $0, -\tfrac{1}{2}, -\tfrac{1}{2}$ & $ 2$ & $ 0$ & $ 2$ & $ 4$ & $ 20$ & $\frac{211}{4}$ & $\frac{35}{4}$ & $-\tfrac{11}{4}$ & $ 0$ &  $\langle s_{15} \rangle $\\
 & $(\rep{3}, \rep{1}, \rep{1}, \rep{1}, \rep{1})$ & $0, -\tfrac{1}{2}, -\tfrac{1}{2}$ & $ 0$ & $\frac{1}{3}$ & $ 0$ & $ 0$ & $ -29$ & $-\tfrac{187}{4}$ & $-\tfrac{11}{4}$ & $-\tfrac{5}{4}$ & $ 0$ &  $\overline{\delta}_{3}$\\
\hline
\hline 
$T_{(0, 1)}^{(n_1, 0, *,*, 0, 0)}$  & $(\rep{1}, \rep{2}, \rep{1}, \rep{1}, \rep{1})$ & $-\tfrac{1}{2}, 0, -\tfrac{1}{2}$ & $-\tfrac{3}{2}$ & $-\tfrac{1}{2}$ & $\frac{27}{2}$ & $ 27$ & $-\tfrac{11}{4}$ & $-\tfrac{13}{4}$ & $\frac{31}{4}$ & $ 1$ & $ 0$ &  $h_{3}, h_{4}$\\
 & $(\rep{1}, \rep{1}, \rep{1}, \rep{1}, \rep{1})$ & $-\tfrac{1}{2}, 0, -\tfrac{1}{2}$ & $\frac{5}{2}$ & $ 0$ & $-\tfrac{25}{2}$ & $ -25$ & $\frac{225}{4}$ & $\frac{33}{4}$ & $\frac{77}{4}$ & $-\tfrac{1}{2}$ & $ 0$ &  $\langle s_{16} , s_{20} \rangle $\\
 & $(\rep{3}, \rep{1}, \rep{1}, \rep{1}, \rep{1})$ & $-\tfrac{1}{2}, 0, -\tfrac{1}{2}$ & $-\tfrac{3}{2}$ & $\frac{1}{3}$ & $\frac{27}{2}$ & $ 27$ & $-\tfrac{11}{4}$ & $-\tfrac{13}{4}$ & $\frac{31}{4}$ & $ 1$ & $ 0$ &  $\overline{\delta}_{4}, \overline{\delta}_{5}$\\
 \hline 
$T_{(0, 1)}^{(n_1, 0, *,* , 0, 1)}$  & $(\rep{1}, \rep{1}, \rep{3}, \rep{1}, \rep{1})$ & $-\tfrac{1}{2}, 0, -\tfrac{1}{2}$ & $\frac{5}{2}$ & $ 0$ & $ -10$ & $\frac{9}{2}$ & $-\tfrac{55}{4}$ & $\frac{59}{2}$ & $-\tfrac{7}{2}$ & $-\tfrac{3}{4}$ & $ 0$ &  $\langle x_{4} , x_{5} \rangle $\\
 & $(\rep{1}, \rep{1}, \rep{1}, \rep{1}, \rep{1})$ & $-\tfrac{1}{2}, 0, -\tfrac{1}{2}$ & $\frac{5}{2}$ & $ 0$ & $ 10$ & $-\tfrac{107}{2}$ & $-\tfrac{55}{4}$ & $\frac{59}{2}$ & $-\tfrac{7}{2}$ & $-\tfrac{3}{4}$ & $ 0$ &  $\langle s_{17} , s_{21} \rangle $\\
 & $(\rep{1}, \rep{1}, \rep{1}, \rep{2}, \rep{1})$ & $-\tfrac{1}{2}, 0, -\tfrac{1}{2}$ & $ 3$ & $ 0$ & $-\tfrac{9}{2}$ & $ -9$ & $-\tfrac{35}{4}$ & $-\tfrac{29}{2}$ & $-\tfrac{7}{2}$ & $-\tfrac{3}{4}$ & $-\tfrac{1}{2}$ &  $\langle z_{3} , z_{5} \rangle $\\
 & $(\rep{1}, \rep{1}, \rep{1}, \rep{1}, \rep{2})$ & $-\tfrac{1}{2}, 0, -\tfrac{1}{2}$ & $ 3$ & $ 0$ & $-\tfrac{9}{2}$ & $ -9$ & $-\tfrac{35}{4}$ & $-\tfrac{29}{2}$ & $-\tfrac{7}{2}$ & $-\tfrac{3}{4}$ & $\frac{1}{2}$ &  $\langle y_{3} , y_{5} \rangle $\\
 \hline 
$T_{(0, 1)}^{(n_1, 0, *, * , 1, 0)}$  & $(\rep{1}, \rep{1}, \rep{1}, \rep{1}, \rep{1})$ & $-\tfrac{1}{2}, 0, -\tfrac{1}{2}$ & $\frac{5}{2}$ & $ 0$ & $\frac{5}{2}$ & $ 5$ & $-\tfrac{45}{4}$ & $-\tfrac{153}{4}$ & $-\tfrac{21}{4}$ & $\frac{7}{2}$ & $ 0$ &  $\langle s_{18} , s_{22} \rangle $\\
 & $(\rep{1}, \rep{1}, \rep{1}, \rep{1}, \rep{1})$ & $-\tfrac{1}{2}, 0, -\tfrac{1}{2}$ & $-\tfrac{3}{2}$ & $ 0$ & $-\tfrac{3}{2}$ & $ -3$ & $\frac{375}{4}$ & $\frac{177}{4}$ & $-\tfrac{43}{4}$ & $ 1$ & $ 0$ &  $\langle s_{19} , s_{23} \rangle $\\
 \hline 
$T_{(0, 1)}^{(n_1, 0, *, * , 1, 1)}$  & $(\rep{1}, \rep{1}, \rep{1}, \rep{1}, \rep{2})$ & $-\tfrac{1}{2}, 0, -\tfrac{1}{2}$ & $ 3$ & $ 0$ & $-\tfrac{9}{2}$ & $ -9$ & $-\tfrac{35}{4}$ & $-\tfrac{29}{2}$ & $-\tfrac{7}{2}$ & $-\tfrac{3}{4}$ & $-\tfrac{1}{2}$ &  $\langle y_{4} , y_{6} \rangle $\\
 & $(\rep{1}, \rep{1}, \rep{1}, \rep{2}, \rep{1})$ & $-\tfrac{1}{2}, 0, -\tfrac{1}{2}$ & $ 3$ & $ 0$ & $-\tfrac{9}{2}$ & $ -9$ & $-\tfrac{35}{4}$ & $-\tfrac{29}{2}$ & $-\tfrac{7}{2}$ & $-\tfrac{3}{4}$ & $\frac{1}{2}$ &  $\langle z_{4} , z_{6} \rangle $  \\
\hline
\hline 
$T_{(1, 1)}^{(n_1, 0, 0, 0, *, *)}$  & $(\rep{1}, \rep{2}, \rep{1}, \rep{1}, \rep{1})$ & $-\tfrac{1}{2}, -\tfrac{1}{2}, 0$ & $\frac{1}{2}$ & $-\tfrac{1}{2}$ & $-\tfrac{29}{2}$ & $ -29$ & $\frac{29}{4}$ & $\frac{1}{4}$ & $-\tfrac{43}{4}$ & $ 1$ & $ 0$ &  $\ell_{1}, \ell_{2}$\\
 & $(\rep{1}, \rep{1}, \rep{1}, \rep{1}, \rep{1})$ & $-\tfrac{1}{2}, -\tfrac{1}{2}, 0$ & $\frac{5}{2}$ & $ 0$ & $\frac{35}{2}$ & $ 35$ & $\frac{265}{4}$ & $\frac{47}{4}$ & $\frac{3}{4}$ & $-\tfrac{1}{2}$ & $ 0$ &  $s_{24}, s_{29}$\\
 & $(\rep{3}, \rep{1}, \rep{1}, \rep{1}, \rep{1})$ & $-\tfrac{1}{2}, -\tfrac{1}{2}, 0$ & $\frac{1}{2}$ & $\frac{1}{3}$ & $-\tfrac{29}{2}$ & $ -29$ & $\frac{29}{4}$ & $\frac{1}{4}$ & $-\tfrac{43}{4}$ & $ 1$ & $ 0$ &  $\overline{d}_{1}, \overline{d}_{2}$\\
 & $(\rep{1}, \rep{1}, \rep{1}, \rep{1}, \rep{1})$ & $-\tfrac{1}{2}, -\tfrac{1}{2}, 0$ & $\frac{3}{2}$ & $ 0$ & $\frac{33}{2}$ & $ 33$ & $-\tfrac{355}{4}$ & $\frac{13}{4}$ & $-\tfrac{31}{4}$ & $ -1$ & $ 0$ &  $\langle s_{25} , s_{30} \rangle $\\
 & $(\rep{1}, \rep{1}, \rep{1}, \rep{1}, \rep{1})$ & $-\tfrac{1}{2}, -\tfrac{1}{2}, 0$ & $-\tfrac{1}{2}$ & $ 0$ & $-\tfrac{31}{2}$ & $ -31$ & $\frac{105}{4}$ & $-\tfrac{9}{4}$ & $-\tfrac{53}{4}$ & $-\tfrac{7}{2}$ & $ 0$ &  $\langle s_{26} , s_{31} \rangle $\\
 \hline 
$T_{(1, 1)}^{(n_1, 0, 0, 1, *, *)}$  & $(\rep{1}, \rep{1}, \rep{1}, \rep{1}, \rep{1})$ & $-\tfrac{1}{2}, -\tfrac{1}{2}, 0$ & $\frac{3}{2}$ & $ 0$ & $ -6$ & $\frac{123}{2}$ & $-\tfrac{75}{4}$ & $ -18$ & $ 15$ & $-\tfrac{3}{4}$ & $ 0$ &  $\langle s_{27} , s_{32} \rangle $\\
 & $(\rep{1}, \rep{1}, \crep{3}, \rep{1}, \rep{1})$ & $-\tfrac{1}{2}, -\tfrac{1}{2}, 0$ & $\frac{3}{2}$ & $ 0$ & $ 14$ & $\frac{7}{2}$ & $-\tfrac{75}{4}$ & $ -18$ & $ 15$ & $-\tfrac{3}{4}$ & $ 0$ &  $\langle\overline{x}_{4}, \overline{x}_{5}\rangle$\\
 \hline 
$T_{(1, 1)}^{(n_1, 0, 1, 0, *, *)}$ & empty &&&&&&&&&&& \\
 \hline 
$T_{(1, 1)}^{(n_1, 0, 1, 1, *, *)}$  & $(\crep{3}, \rep{2}, \rep{1}, \rep{1}, \rep{1})$ & $-\tfrac{1}{2}, -\tfrac{1}{2}, 0$ & $\frac{3}{2}$ & $\frac{1}{6}$ & $\frac{3}{2}$ & $ 3$ & $\frac{147}{4}$ & $ 6$ & $ -5$ & $\frac{1}{4}$ & $ 0$ &  $q_{1}, q_{2}$\\
 & $(\rep{3}, \rep{1}, \rep{1}, \rep{1}, \rep{1})$ & $-\tfrac{1}{2}, -\tfrac{1}{2}, 0$ & $\frac{3}{2}$ & $-\tfrac{2}{3}$ & $\frac{3}{2}$ & $ 3$ & $\frac{147}{4}$ & $ 6$ & $ -5$ & $\frac{1}{4}$ & $ 0$ &  $\overline{u}_{1}, \overline{u}_{2}$\\
 & $(\rep{1}, \rep{1}, \rep{1}, \rep{1}, \rep{1})$ & $-\tfrac{1}{2}, -\tfrac{1}{2}, 0$ & $\frac{3}{2}$ & $ 1$ & $\frac{3}{2}$ & $ 3$ & $\frac{147}{4}$ & $ 6$ & $ -5$ & $\frac{1}{4}$ & $ 0$ &  $\overline{e}_{1}, \overline{e}_{2}$\\
 & $(\rep{1}, \rep{2}, \rep{1}, \rep{1}, \rep{1})$ & $-\tfrac{1}{2}, -\tfrac{1}{2}, 0$ & $\frac{3}{2}$ & $\frac{1}{2}$ & $\frac{3}{2}$ & $ 3$ & $-\tfrac{259}{4}$ & $\frac{5}{2}$ & $\frac{27}{2}$ & $\frac{1}{4}$ & $ 0$ &  $\overline{h}_{3}, \overline{h}_{4}$\\
 & $(\crep{3}, \rep{1}, \rep{1}, \rep{1}, \rep{1})$ & $-\tfrac{1}{2}, -\tfrac{1}{2}, 0$ & $\frac{3}{2}$ & $-\tfrac{1}{3}$ & $\frac{3}{2}$ & $ 3$ & $-\tfrac{259}{4}$ & $\frac{5}{2}$ & $\frac{27}{2}$ & $\frac{1}{4}$ & $ 0$ &  $\delta_{4}, \delta_{5}$\\
 & $(\rep{1}, \rep{1}, \rep{1}, \rep{1}, \rep{1})$ & $-\tfrac{1}{2}, -\tfrac{1}{2}, 0$ & $-\tfrac{9}{2}$ & $ 0$ & $-\tfrac{9}{2}$ & $ -9$ & $-\tfrac{35}{4}$ & $-\tfrac{29}{2}$ & $-\tfrac{7}{2}$ & $-\tfrac{3}{4}$ & $ 0$ &  $\langle s_{28} , s_{33} \rangle $\\
\end{longtable}
}

\section{Selection rules}
\label{app:SelectionRules}

The usual string selection rules are modified due to the freely acting symmetry
which we mod out. Starting from the general rules
\cite{Hamidi:1986vh,Dixon:1986qv} we find that a superpotential term $\prod_i
\Phi^{(i)}$ between the superfields $\Phi^{(i)}$ is allowed if the following
conditions are met:
\begin{subequations}
\begin{eqnarray}
\text{gauge invariance} & : & \sum_i p_\mathrm{sh}^{(i)} ~=~ 0\;,\\
\text{$R$-invariance}   & : & \sum_i R^{(i)}    ~=~ (-1,-1,-1) \mod (2,2,2)\;,\\
\text{point group rule} & : & \sum_i k^{(i)}    ~=~ 0 \mod 2\;,\\
                        &   & \sum_i \ell^{(i)} ~=~ 0 \mod 2\;,\\
\text{space group rule} & : & \sum_i n_1^{(i)}  ~=~ 0 \mod 2\;, \label{rule1}\\
                        &   & \sum_i n_3^{(i)}  ~=~ 0 \mod 2\;,\\
                        &   & \sum_i n_5^{(i)}  ~=~ 0 \mod 2\;, \label{rule5}\\
                        &   & \sum_i (n_2^{(i)} + n_4^{(i)} + n_6^{(i)}) ~=~ 0 \mod 2\;.
\label{newrule} 
\end{eqnarray}
\end{subequations}
Here $p_\mathrm{sh}^{(i)}$ denote the shifted $\E8\times\E8$ momenta, the
discrete $R$ charges $R^{(i)}$ are computed from the (shifted) $\SO8$ momenta
and oscillator quantum numbers, $R^{(j)} = q_\mathrm{sh}^{(j)} -
\widetilde{N}^{(j)} + \tilde{N}^{*\,(j)}$, $k^{(i)}$, $\ell^{(i)}$ and
$n_\alpha^{(i)}$ specify the constructing element
$(\theta^{k^{(i)}}\,\omega^{\ell^{(i)}},n_\alpha^{(i)}\,e_\alpha)$ of the
corresponding state. In contrast to the space group selection rule of the
standard $\Z2\times\Z2$ orbifold, where similar conditions to the rules
\eqref{rule1} to \eqref{rule5} also apply for $n_{\alpha}^{(i)}$, $\alpha =
2,4,6$, we find the single condition \eqref{newrule} in the freely acting case. 

\enlargethispage{1cm}
\bibliography{Orbifold}
\bibliographystyle{ArXiv}

\end{document}